\documentstyle[aps,preprint,pra,url]{revtex}
\begin{document}
\newcommand{\etal}{{\em et al.}\/}
\newcommand{\IP}{inner polarization}
\newcommand{\IPF}{\IP\ function}
\newcommand{\IPFs}{\IP\ functions}
\newcommand{\auth}[2]{#1 #2, }
\newcommand{\jcite}[4]{#1 {\bf #2}, #3 (#4)}
\newcommand{\et}{ and }
\newcommand{\twoauth}[4]{#1 #2 and #3 #4,}
\newcommand{\oneauth}[2]{#1 #2,}
\newcommand{\andauth}[2]{and #1 #2, }
\newcommand{\book}[4]{{\it #1} (#2, #3, #4)}
\newcommand{\erratum}[3]{\jcite{erratum}{#1}{#2}{#3}}
\newcommand{\inpress}[1]{{\it #1}}
\newcommand{\inbook}[5]{In {\it #1}; #2; #3: #4, #5}
\newcommand{\JCP}[3]{\jcite{J. Chem. Phys.}{#1}{#2}{#3}}
\newcommand{\jms}[3]{\jcite{J. Mol. Spectrosc.}{#1}{#2}{#3}}
\newcommand{\jmsp}[3]{\jcite{J. Mol. Spectrosc.}{#1}{#2}{#3}}
\newcommand{\jmstr}[3]{\jcite{J. Mol. Struct.}{#1}{#2}{#3}}
\newcommand{\cpl}[3]{\jcite{Chem. Phys. Lett.}{#1}{#2}{#3}}
\newcommand{\cp}[3]{\jcite{Chem. Phys.}{#1}{#2}{#3}}
\newcommand{\pr}[3]{\jcite{Phys. Rev.}{#1}{#2}{#3}}
\newcommand{\jpc}[3]{\jcite{J. Phys. Chem.}{#1}{#2}{#3}}
\newcommand{\jpcA}[3]{\jcite{J. Phys. Chem. A}{#1}{#2}{#3}}
\newcommand{\jpca}[3]{\jcite{J. Phys. Chem. A}{#1}{#2}{#3}}
\newcommand{\jpcB}[3]{\jcite{J. Phys. Chem. B}{#1}{#2}{#3}}
\newcommand{\jpB}[3]{\jcite{J. Phys. B}{#1}{#2}{#3}}
\newcommand{\PRA}[3]{\jcite{Phys. Rev. A}{#1}{#2}{#3}}
\newcommand{\PRB}[3]{\jcite{Phys. Rev. B}{#1}{#2}{#3}}
\newcommand{\PRL}[3]{\jcite{Phys. Rev. Lett.}{#1}{#2}{#3}}
\newcommand{\jcc}[3]{\jcite{J. Comput. Chem.}{#1}{#2}{#3}}
\newcommand{\molphys}[3]{\jcite{Mol. Phys.}{#1}{#2}{#3}}
\newcommand{\mph}[3]{\jcite{Mol. Phys.}{#1}{#2}{#3}}
\newcommand{\APJ}[3]{\jcite{Astrophys. J.}{#1}{#2}{#3}}
\newcommand{\cpc}[3]{\jcite{Comput. Phys. Commun.}{#1}{#2}{#3}}
\newcommand{\jcsfii}[3]{\jcite{J. Chem. Soc. Faraday Trans. II}{#1}{#2}{#3}}
\newcommand{\FD}[3]{\jcite{Faraday Discuss.}{#1}{#2}{#3}}
\newcommand{\prsa}[3]{\jcite{Proc. Royal Soc. A}{#1}{#2}{#3}}
\newcommand{\jacs}[3]{\jcite{J. Am. Chem. Soc.}{#1}{#2}{#3}}
\newcommand{\joptsa}[3]{\jcite{J. Opt. Soc. Am.}{#1}{#2}{#3}}
\newcommand{\cjc}[3]{\jcite{Can. J. Chem.}{#1}{#2}{#3}}
\newcommand{\ijqcs}[3]{\jcite{Int. J. Quantum Chem. Symp.}{#1}{#2}{#3}}
\newcommand{\ijqc}[3]{\jcite{Int. J. Quantum Chem.}{#1}{#2}{#3}}
\newcommand{\spa}[3]{\jcite{Spectrochim. Acta A}{#1}{#2}{#3}}
\newcommand{\tca}[3]{\jcite{Theor. Chem. Acc.}{#1}{#2}{#3}}
\newcommand{\tcaold}[3]{\jcite{Theor. Chim. Acta}{#1}{#2}{#3}}
\newcommand{\jpcrd}[3]{\jcite{J. Phys. Chem. Ref. Data}{#1}{#2}{#3}}
\newcommand{\science}[3]{\jcite{Science}{#1}{#2}{#3}}
\newcommand{\CR}[3]{\jcite{Chem. Rev.}{#1}{#2}{#3}}
\newcommand{\bbpc}[3]{\jcite{Ber. Bunsenges. Phys. Chem.}{#1}{#2}{#3}}
\newcommand{\acie}[3]{\jcite{Angew. Chem. Int. Ed.}{#1}{#2}{#3}}
\newcommand{\ijck}[3]{\jcite{Int. J. Chem. Kinet.}{#1}{#2}{#3}}
\newcommand{\jct}[3]{\jcite{J. Chem. Thermodyn.}{#1}{#2}{#3}}

\newcommand{\deltah}[0]{$\Delta$H}
\newcommand{\deltahf}[0]{$\Delta$H$_f$}
\newcommand{\hoof}[0]{$\Delta$H$_f$$^{298}$}
\newcommand{\hof}[0]{$\Delta H^\circ_f$}
\newcommand{\hofzero}[0]{$\Delta$H$_f$$^{0}$}
\newcommand{\abin}{{\em ab initio}}

\draft
\title{Fully ab initio atomization energy of benzene via W2 theory}
\author{Srinivasan Parthiban and Jan M.L. Martin*}
\address{Department of Organic Chemistry,
Kimmelman Building, Room 262,
Weizmann Institute of Science,
IL-76100 Re\d{h}ovot, Israel. {\rm E-mail:} {\tt comartin@wicc.weizmann.ac.il}
}
\date{{\em J. Chem. Phys.} N1.03.005 received March 9, 2001; accepted May 18, 2001}
\maketitle
\begin{abstract}
The total atomization energy at absolute zero, (TAE$_0$) of benzene, C$_6$H$_6$, was computed fully
{\em ab initio} by means of W2h theory as 1306.6 kcal/mol, to be
compared with the experimentally derived value 1305.7$\pm$0.7 kcal/mol.
The computed result includes contributions from inner-shell correlation
(7.1 kcal/mol), scalar relativistic effects (-1.0 kcal/mol), atomic
spin-orbit splitting (-0.5 kcal/mol), and the anharmonic zero-point
vibrational energy (62.1 kcal/mol). The largest-scale calculations involved are
CCSD/cc-pV5Z and CCSD(T)/cc-pVQZ; basis set extrapolations account for
6.3 kcal/mol of the final result. Performance of more approximate
methods has been analyzed. Our results suggest that, even for systems
the size of benzene, chemically accurate
molecular atomization energies can be obtained from fully first-principles
calculations, without resorting to corrections or parameters derived from experiment.
\end{abstract}
\clearpage
Computational thermochemistry is coming of age as part of the chemist's toolbox\cite{acs677}.
Popular approaches (such as G3 theory\cite{g3paper} and CBS-QB3\cite{cbs3}) that can lay
claim to `chemical accuracy' (1 kcal/mol) on average for small systems, invariably
rely on a combination of relatively low-level {\em ab initio} calculations and sophisticated
empirical correction schemes, which have been parametrized against experimental data.

In recent years, a number of groups have focused on obtaining accurate 
thermodynamic data of small
molecules by means of fully {\em ab initio} approaches (i.e. devoid of parameters derived from
experiment); the reader is referred to studies by e.g. Dixon\cite{Fel99,Dix2000c6h6}, 
Klopper\cite{KlopRev}, Bauschlicher\cite{Bau98}, and Martin\cite{nato}. Very recently,
we developed two near-black box methods of this type,
known as W1 and W2 theory (for Weizmann-1 and -2, respectively);
in the original paper\cite{w1} and a subsequent validation study\cite{w1w2validate} for most of the
G2/97 data set\cite{g2hof,g2ipea}, we have shown that these methods
yield thermochemical 
data in the kJ/mol accuracy range for small systems that are well described by a single reference
configuration. 

The question arises as to how well such methods would `scale up' to larger systems. For
this purpose, the ubiquitous benzene molecule would appear to offer an excellent
`stress test'. It has six heavy atoms, yet its heat of formation is known precisely
from experiment, and its high symmetry makes it amenable to fairly large-scale
treatments with modern high-performance computing hardware. In the present note,
we shall discuss the performance the total atomization energy (TAE$_e$ if zero-point exclusive,
TAE$_0$ at 0 K) of benzene of the more rigorous W2h theory, of the more
widely applicable W1 and W1h theories, and of a variety of more approximate approaches.

All calculations involved in W1, W1h, and W2h theory were carried out using 
MOLPRO 98.1\cite{molpro98} running on a Compaq ES40 minisupercomputer in our
laboratory. (For the open-shell calculations on carbon, the definition of the
CCSD(T)\cite{Rag89} energy according to Ref.\cite{Wat93} has been used.)
Detailed descriptions and justifications of the various steps involved
can be found in Refs.\cite{w1,w1w2validate}. We merely note here that, for the 
system under study, the 
final result at the highest level of theory (W2h) 
consists of the following components: (a) an SCF limit
extrapolated from SCF/cc-pV$n$Z (correlation consistent polarized valence $n$-tuple zeta\cite{Dun89},
with $n$=T,Q,5) energies using the formulas $E(n)=E_\infty+B/C^n$ (old style\cite{w1})
or $E(n)=E_\infty+A/n^5$ (new style\cite{w1w2validate}); (b) a CCSD valence correlation limit extrapolated
from CCSD/cc-pV$n$Z ($n$=Q,5) results using $E(n)=E_\infty+A/n^3$; (c) a limit for the effect
of connected triple excitations extrapolated from [CCSD(T)/cc-pV$n$Z--CCSD/cc-pV$n$Z] ($n$=T,Q) using
$E(n)=E_\infty+A/n^3$; (d) an inner-shell correlation contribution obtained at the 
CCSD(T)/MTsmall level; (e) a scalar relativistic (1st-order Darwin and mass-velocity\cite{Cow76,Mar83}) contribution
obtained as an expectation value from the ACPF/MTsmall\cite{Gda88} wave function; (f) a first-order
spin-orbit correction derived from the fine structure of the constituent atoms; and 
(g) the anharmonic zero-point energy (vide infra). The computationally most intensive
step was the CCSD/cc-pV5Z calculation. At 876 basis functions, with 30 electrons
correlated, this could not be carried out using a conventional algorithm even
while exploiting the $D_{2h}$ subgroup of $D_{6h}$; using the direct CCSD algorithm
of Lindh, Sch\"utz, and Werner\cite{dirccsd} as implemented in MOLPRO, it took
14 days of CPU time on a 667 MHz Alpha EV67 CPU with 768 MB of memory
allocated. (The CCSD(T)/cc-pVQZ optimum geometry required for the 
W2h calculations was taken from Ref.\cite{StaGau2000}.) 

The W1h calculations primarily differ in that the extrapolations are carried out
with smaller cc-pV$n$Z ($n$=D,T,Q) basis sets (and $E(n)=E_\infty+A/n^{3.22}$ for
the correlation steps, see\cite{w1} for its derivation), while in W1 theory,
the carbon basis set is in addition augmented with diffuse functions\cite{Ken92}.
All relevant data for the W2h calculation are collected in Table \ref{tab:c6h6}.
Calculations using more approximate methods such as G3 theory\cite{g3paper} and
CBS-QB3\cite{cbs3} were carried out using their respective implementations in Gaussian 98\cite{g98revA7}.

For a molecule this size, the zero-point vibrational energy (ZPVE) is large enough that 
even fairly small relative errors may compromise the quality of the final TAE.
Handy and coworkers\cite{HandyC6H6} computed a quartic force field at the
B3LYP/TZ2P\cite{Bec93,LYP} level; from their published anharmonicity constants
(in particular the set deperturbed for Fermi resonances closer than 100 cm$^{-1}$),
we obtain an anharmonic ZPVE of 62.04 kcal/mol. At the same level of theory, one-half the sum of the
harmonics, $\sum_i{\omega_id_i/2}$ (with $d_i$ the degeneracy of mode $i$) 
comes out 0.9 kcal/mol too high at 62.96 kcal/mol, while 
one-half the sum of the fundamentals, $\sum_i{\nu_id_i/2}$, comes out 1 kcal/mol too low at 
60.98 kcal/mol. The average of both estimates, $\sum_i{(\omega_i+\nu_i)d_i/4}$=61.97 kcal/mol, is only 
0.07 kcal/mol below the true anharmonic value. 
From the best available computed harmonic frequencies, CCSD(T)/ANO4321\cite{c6h6}
and the best available experimental fundamentals\cite{HandyC6H6}, we obtain ZPVE=62.01 kcal/mol,
or, after correction for the difference at the B3LYP/TZ2P level between $\sum_i{(\omega_i+\nu_i)d_i/4}$
and the true anharmonic ZPVE, we find a best-estimate ZPVE=62.08 kcal/mol.

Of the more approximate approaches used in  various computational thermochemistry methods,
HF/6-31G* harmonic frequencies scaled by 0.8929 (as used in G2 and G3 theory\cite{g3paper}) yield 60.33 kcal/mol,
or about 1.7 kcal/mol too low. The procedure used in the very recent G3X and G3SX theories\cite{g3x},
B3LYP/6-31G(2df,p) scaled by 0.9854, however reproduces the best estimate to within 0.1 kcal/mol.
B3LYP/6-311G** harmonic frequencies scaled by 0.99, as used
in CBS-QB3\cite{cbs3}, yields 62.23 kcal/mol, in very good agreement with the best estimate;
the HF/6-31G(d) scaled by 0.9184 estimate in CBS-Q yields 61.69 kcal/mol, slightly too low.
Finally, B3LYP/cc-pVTZ harmonics scaled by 0.985 (as used in W1 and W1h theory\cite{w1}) yield
62.04 kcal/mol, in near-perfect agreement with the best estimate.

Relevant data for the W2h calculation are collected in Table \ref{tab:c6h6}.
At first sight, the disagreement between the W2h $\Delta H^\circ_{f, 0 K}$=23.1 kcal/mol
and the experimental value of 24.0$\pm$0.2 kcal/mol seems disheartening for such a CPU-intensive
calculation. 
(Note that it `errs' on the far side of the most recent previous benchmark calculation\cite{Dix2000c6h6},
24.7$\pm$0.3 kcal/mol, which used similar-sized basis sets as W1 theory.)
However, the comparison with experiment
is not entirely `fair' since it neglects the experimental uncertainties in the atomic heats of formation
required to convert an atomization energy into a heat of formation (or vice versa). Combining these
with the experimental $\Delta H^\circ_{f, 0 K}$ leads to an experimentally derived 
TAE$_0$=1305.7$\pm$0.7 kcal/mol, where the uncertainty is dominated by six times that in the
heat of vaporization of graphite. In other words, our calculated TAE$_0$=1306.8 kcal/mol is
only 0.2 kcal/mol removed from the upper end of the experimental uncertainty interval.
(After all, an error of 0.02\% seems to be a bit much to ask for.)  

Alternatively and equivalently, one could affix an uncertainty of $\pm$0.7
kcal/mol to the computed W2h $\Delta H^\circ_{f, 0 K}$=23.1$\pm$0.7 kcal/mol,
where the error bar only reflects the uncertainties in the auxiliary 
experimental data (i.e. the heats of atomization of the elements), but
does not include the uncertainty in the theoretical calculation itself
which is harder to quantify. While most chemists would prefer the
heat of formation, an analysis in terms of atomization energies is somewhat
more elegant since it avoids mixing computed and observed data. 
(Unfortunately, a benchmark {\em ab initio} heat of vaporization of graphite
does not appear to be feasible at this point in time.)

Secondly, let us consider the `gaps' bridged by the extrapolations. For the SCF component,
that is a very reasonable 0.3 kcal/mol (0.03 \%), but for the CCSD valence correlation
component this rises to 5 kcal/mol (1.7 \%) while for the connected triple excitations
contribution it amounts to 1 kcal/mol (3.7 \% --- note however that a smaller basis set
is being used than for CCSD). It is clear that the extrapolations are indispensable to
obtain even a {\em useful} result, let alone an accurate one, even with such large basis
sets. 

Inner-shell correlation, at 7 kcal/mol, is of quite nontrivial importance, but even scalar relativistic
effects (at $-$1.0 kcal/mol) cannot be ignored. (The discrepancy between our scalar relativistic
correction and the previous SCF-level calculation of Kedziora et al.\cite{Ked99}, $-$1.27 kcal/mol, is 
consistent with the known tendency\cite{w1,msft,Bau2000} of SCF-level scalar relativistic 
corrections to be overestimated by 20--25\%.) 
And manifestly, even a 2\% error in a 62 kcal/mol zero-point
vibrational energy would be unacceptable. 

Let us now consider the more approximate results. While W1h coincidentally agrees to better than 0.2 
kcal/mol with the W2h result, W1 deviates from the latter by 0.6 kcal/mol. Note however that in 
W1h theory, the extrapolations bridge gaps of 0.8 (SCF), 10.1 (CCSD), and 2.1 (T) kcal/mol,
the corresponding amounts for W1 theory being 0.7, 9.1, and 1.9 kcal/mol, respectively. 
Common sense suggests that if extrapolations account for 13.0 (W1h) and 11.7 (W1) kcal/mol, then
a discrepancy of 1 kcal/mol should not come as a surprise --- in fact, the relatively good agreement
between the two sets of numbers and the more rigorous W2h result (total extrapolation: 6.3 kcal/mol)
testifies, if anything, to the robustness of the method.

As for the difference of about 0.4 kcal/mol between the old-style\cite{w1} and new-style\cite{w1w2validate} SCF 
extrapolations in W1h and W1 theories, comparison with the W2h SCF limits clearly confirms 
the new-style extrapolation to be the more reliable one. (The two extrapolations yield basically
the same result in W2h.) This should not be seen as an indication that the $E_\infty+A/L^5$ formula
is somehow better founded theoretically, but rather as an example of why reliance on (aug-)cc-pVDZ
data should be avoided if at all possible. 

Our best TAE$_0$ value (W2h) differs by 1.6 kcal/mol from the previous 
benchmark calculation
of Feller and Dixon\cite{Dix2000c6h6}. In fact, since their largest basis set
is of AVQZ quality, the appropriate comparison would 
be with our W1 atomization energy, which is 2.3 kcal/mol larger
than their result using RCCSD(T) atomic energies. 
The zero-point energy and the corrections
for core correlation, scalar relativistic effects, and atomic 
spin-orbit splitting are all very similar in the two studies. Their 
extrapolation approach is very different from ours, but in the event this
difference nearly cancels out with that caused by the different definitions
of the RCCSD(T) energy used in the atomic calculations. (Feller and
Dixon followed Ref.\cite{Ham93}, as opposed to Ref.\cite{Wat93} in the
present paper: 
we find the difference for six carbon atoms to be 0.52 kcal/mol
at the CCSD(T)/AVQZ level.) The difference is in fact mostly due to
a $-$2.1 kcal/mol correction for `higher-order correlation effects'
applied in Ref.\cite{Dix2000c6h6}, which is an estimate of the
CCSDT $-$ CCSD(T) difference from small basis set calculations. 
However, the generally excellent quality of CCSD(T) computed 
bond energies rests to a large extent on an error compensation between
neglect of higher-order connected triple excitations (which tend
to reduce the binding energy) and complete neglect of quadruple 
excitations (which tend to increase it) \cite{Bak2000}.
It has been known for some time (e.g.\cite{Wat90}) that CCSDT energies
are not necessarily closer to full CI than CCSD(T). Consequently, an
accurate treatment should either include both $T_4$ and higher-order
$T_3$ effects where it is possible to do so, or neglect both: including
only the higher-order $T_3$ of necessity leads to an underestimate
of TAE. We do note that our respective best estimates bracket the experimental
value, which may indicate that the `true' (full CI) TAE lies in between.
However, in view of the uncertainty on the experimental TAE
and the impossibility to carry out even a highly approximate CCSDTQ calculation
on benzene, it is hard to make a definite statement
about this.

Turning finally to the more approximate approaches, 
G2 theory clearly underestimates TAE$_0$: 
G3 represents a major improvement, but the better than 1 kcal/mol agreement between the
G3 TAE$_0$ and the experimentally derived value in fact benefits from an error compensation
with the underestimated ZPVE: a rather more pronounced difference is seen for TAE$_e$. 
This problem is remedied in the very recent G3X and G3SX theories, which predict both
TAE$_e$ and TAE$_0$ to within 1 kcal/mol of experiment, as does CBS-QB3. CBS-Q is
slightly too low; the fairly elaborate CBS-APNO method\cite{Pet94} find results
that nearly coincide with W1 theory. (We note that none of the G$n$ and CBS methods considered 
explicitly includes scalar relativistic effects; they instead rely on them being absorbed into the
parametrization.)

Summarizing the above, we may state the following:

The total atomization energy of benzene, C$_6$H$_6$, was computed fully
{\em ab initio} by means of W2h theory as 1306.6 kcal/mol, to be 
compared with the experimentally derived value 1305.7$\pm$0.7 kcal/mol.
The computed result includes contributions from inner-shell correlation
(7.1 kcal/mol), scalar relativistic effects (-1.0 kcal/mol), atomic
spin-orbit splitting (-0.5 kcal/mol), and the anharmonic zero-point
vibrational energy (62.1 kcal/mol). The largest-scale calculations involved are
CCSD/cc-pV5Z and CCSD(T)/cc-pVQZ; basis set extrapolations account for
6.3 kcal/mol of the final result. Performance of more approximate
methods has been analyzed. Our results suggest that, even for systems
the size of benzene, chemically accurate
molecular atomization energies can be obtained from fully first-principles
calculations, without resorting to corrections or parameters derived from experiment.

\acknowledgments
SP acknowledges a Postdoctoral Fellowship from the Feinberg 
Graduate School (Weizmann Institute). JM is the incumbent of the Helen 
and Milton A. Kimmelman Career Development Chair. This research was
supported by the 
{\em Tashtiyot} Program of the Ministry of Science (Israel).

\begin{table}[h]
\caption{Individual components in W1h, W1, and W2h total atomization energy
{\em cum} heat of formation of benzene. All data in kcal/mol\label{tab:c6h6}}
\squeezetable
\begin{tabular}{lrrrrrr}
Ref. geom.    & \multicolumn{4}{c}{B3LYP/cc-pVTZ} & \multicolumn{2}{r}{CCSD(T)/cc-pVQZ}\\
              & \multicolumn{2}{c}{W1h} & \multicolumn{2}{c}{W1} & \multicolumn{2}{c}{W2h}\\
\hline
SCF           & VDZ & 1024.19 & A$'$VDZ & 1024.59 & VTZ & 1042.16 \\
              & VTZ & 1042.10 & A$'$VTZ & 1042.62 & VQZ & 1044.62 \\
              & VQZ & 1044.56 & A$'$VQZ & 1044.84 & V5Z & 1045.30 \\
old-style $E_\infty+A/B^n$ & V$\infty$Z & 1044.95 & V$\infty$Z & 1045.15 & V$\infty$Z & 1045.56 \\
new-style $E_\infty+A/n^5$ & V$\infty$Z & 1045.33 & V$\infty$Z & 1045.53 & V$\infty$Z & 1045.63 \\
CCSD          & VDZ &  225.94 & A$'$VDZ &  226.11 & VTZ &  265.49\\
              & VTZ &  265.55 & A$'$VTZ &  268.44 & VQZ &  280.91\\
              & VQZ &  280.97 & A$'$VQZ &  282.39 & V5Z &  285.72\\
       & V$\infty$Z &  291.08 & V$\infty$Z &  291.53 & V$\infty$Z & 290.77 \\
(T)           & VDZ &   18.72 & A$'$VDZ &   19.64 & VTZ &   24.41\\
              & VTZ &   24.42 & A$'$VTZ &   24.78 & VQZ &   25.74\\
       & V$\infty$Z &   26.55 & V$\infty$Z &   26.69 & V$\infty$Z &  26.71 \\
\multicolumn{2}{l}{Inner-shell correlation}  & 7.09 &  & 7.08 &  & 7.10 \\
\multicolumn{2}{l}{Darwin and mass-velocity} & -0.99 &  & -0.99 &  & -0.99 \\
\multicolumn{2}{l}{Spin-orbit coupling} & -0.51 &  & -0.51 &  & -0.51 \\
\hline
        &   TAE$_e$ & ZPVE    & TAE$_0$ & $\Delta H^\circ_{f, 0 K}$ & $H_{298}-H_0$ & $\Delta H^\circ_{f,298 K}$\\
\hline
Expt.   &1367.8$\pm$0.7$^b$ & 62.08$^a$& 1305.7$\pm$0.7$^b$ & 24.0$\pm$0.12 & 3.43 & 19.82$\pm$0.12\\
W2h     & 1368.71   & 62.08$^a$& 1306.63  &   23.01  &  3.34  &  18.78 \\ 
W1      & 1369.33   & 62.04   &  1307.29  &   22.39  &  3.34  &  18.15 \\
W1h     & 1368.54   & 62.04   &  1306.49  &   23.18  &  3.34  &  18.95 \\
G3X     & 1367.13   & 61.93   &  1305.20  &   24.5   &  3.42  &  20.3\\
G3SX    & 1366.92   & 61.93   &  1304.99  &   24.7   &  3.42  &  20.5\\
G3      & 1365.48   & 60.33   &  1305.15  &   24.5   &  3.39  &  20.4\\
G2$^c$  & 1362.24   & 60.33   &  1301.91  &   27.8   &  3.39  &  23.7\\
CBS-QB3  & 1365.94   & 62.23   &  1303.71  &   25.95  &  3.72  &  22.11\\
CBS-Q    & 1365.45   & 61.69   &  1303.76  &   25.90  &  3.53  &  21.87\\
CBS-APNO & 1369.34   & 61.88   &  1307.46  &   22.20  &  3.53  &  18.17\\
\end{tabular}

(a) best estimate (see text).\\
(b) From $\Delta H^\circ_{f, 0 K}$[C$_6$H$_6$(g)]=24.0$\pm$0.12 kcal/mol\cite{webbook,trc},
$\Delta H^\circ_{f,0}$[C(g)]=169.98$\pm$0.11 kcal/mol\cite{Cod89}, and
$\Delta H^\circ_{f,0}$[H(g)]=51.634 kcal/mol\cite{Cod89}.
(The uncertainty in $\Delta H^\circ_{f,0}$[H(g)] is negligible.)\\
(c) All values except G2 include corrections for atomic spin-orbit splitting.\\
\typeout{Note S-O correction in Gaussian 98 only automatic for atoms in G3 theory variants}
\end{table}
\end{document}